\documentclass[10pt,letterpaper]{article}
\usepackage[top=0.85in,footskip=0.75in,marginparwidth=2in]{geometry}

\usepackage[utf8]{inputenc}

\usepackage{cite}

\usepackage{nameref,hyperref}

\usepackage{microtype}
\DisableLigatures[f]{encoding = *, family = * }

\usepackage{changepage}

\usepackage[aboveskip=1pt,labelfont=bf,labelsep=period,singlelinecheck=off]{caption}

\makeatletter
\renewcommand{\@biblabel}[1]{\quad#1.}
\makeatother

\usepackage{lastpage,fancyhdr,graphicx}
\usepackage{epstopdf}
\fancyheadoffset[L]{2.25in}
\fancyfootoffset[L]{2.25in}

\usepackage{graphicx}  % needed for figures
\usepackage{dcolumn}   % needed for some tables
\usepackage{bm}        % for math
\usepackage{amssymb}   % for math
\usepackage{amsmath}
\usepackage{mciteplus}
\usepackage{color}

\definecolor{Gray}{gray}{.25}

\usepackage{graphicx}

\usepackage{sidecap}

\usepackage{wrapfig}
\usepackage[pscoord]{eso-pic}
\usepackage[fulladjust]{marginnote}
\reversemarginpar

\begin{document}
\vspace*{0.35in}

% title goes here:
\begin{flushleft}
{\Large
\textbf\newline{Corrected thermodynamics of nonlinear magnetic-charged black hole surrounded by perfect fluid dark matter}
}
\newline
% authors go here:
\\
Ragil Brand Tsafack Ndongmo\textsuperscript{1,*},
Saleh Mahamat\textsuperscript{2,3},
Thomas Bouetou Bouetou\textsuperscript{1,4},
Conrad Bertrand Tabi\textsuperscript{5},
Timoleon Crepin Kofane\textsuperscript{1,5}
\\
\bigskip
\bf{1} Department of Physics, Faculty of Science, University of Yaounde I, P.O. Box. 812, Yaounde, Cameroon,
\\

\bf{2} Department of Physics, Higher Teacher’s Training College,  University of Maroua, P.O. Box 55, Maroua, Cameroon,
\\

\bf{3}Centre de Recherche en Didactique des Sciences Fondamentales et Appliquées (CERDISFA), University of Maroua, P.O. Box 55 Maroua, Cameroon,
\\

\bf{4} National Advanced School of Engineering, University of Yaounde I, P.O. Box. 8390, Yaounde, Cameroon,
\\
\bf{5} Department of Physics and Astronomy, Botswana International University of Science and Technology, Private Mail Bag 16, Palapye, Botswana
\\
\bigskip
* nragilbrand@gmail.com

\end{flushleft}

\begin{abstract}
	
		In this paper, we investigate the influence of perfect fluid dark matter and quantum corrections on the thermodynamics of nonlinear magnetic-charged black hole. We consider the metric of the static nonlinear magnetic-charged black hole in the background of perfect fluid dark matter. Starting with the black hole temperature and the corrected entropy, we use the event horizon propriety in order to find the temperature, and based on the surface gravity definition, we find the uncorrected entropy. However, using the definition of the corrected entropy due to thermal fluctuation, we find and plot the entropy of the black hole. We find that the entropy is affected for smaller nonlinear magnetic-charged black holes. Afterwards, we study the thermodynamic stability of the black hole by computing and plotting the evolution of heat capacity. The results show that second-order phase transition occurs, which appears more later as the dark matter parameter decreases, and leads the black hole to move from the stable phase to the unstable phase. Furthermore, we show that the heat capacity for smaller black holes are also affected, since it appears not being only an increasing function. We also find that the behavior of Gibbs energy is modified when taking into account quantum corrections.
	
	%	 for the thermodynamic analysis of rotating and nonlinear magnetic-charged black hole with quintessence. Accordingly, we compute various thermodynamics quantities of the black hole, such as mass, temperature, potential provided from the magnetic charge, and the heat capacity. Moreover, we study phase transitions of this black hole, analyzing the plot of its heat capacity. Then, we have shown that the black hole mass would have a phase of decrease, while the temperature increases from negative absolute temperatures. From the behavior of the heat capacity, we point out that the black hole undergoes to a second-order phase transition, which is shifted towards the higher values of entropy as we increase the rotating parameter $a$ or the magnetic parameter $Q$.

	%We purpose a new approach of black hole mass decreasing, which takes into account thermodynamics and hydrodynamics processes, in the presence of quintessence and phantom dark energies. Accordingly, we insert some terms into the Schwarzschild metric in order to obtain a new expression of the black hole mass. Further, we show that by the thermodynamics process, the second-order phase transition does not occur when we take into account phantom energy, except for complex entropies or relativistic time. At the end, we show a new principle to analyze black holes coalescence into a space-time, dilated by dark energy.
\end{abstract}

\section{Introduction}

Black holes represent one of the most fascinating objects studied in astrophysics and cosmology.  The mathematical framework necessary to study them is the General theory of relativity, established by Einstein\cite{einstein1916The}. The interaction between black holes and their surrounds could lead to have an electromagnetic distribution around the event horizon of black holes. On that way, the first solution of Einstein field equations coupled to Maxwell equations was found by Hans Reissner\cite{Reissner1916uber} by Gunnar Nordst\"om\cite{nordstrom1918on}. However, because of the problems of singularity of an electric field in the center of particles and the infinite electromagnetic energy in classical electrodynamics, strong electromagnetic fields has been considered. Hence, several models of nonlinear electrodynamics have been proposed, which have at weak-field limit Maxwell's electrodynamics (NED) \cite{kruglov2015model,kruglov2015nonlinear,kruglov2017notes,bandos2020nonlinear,born1934quantum,kosyakov2020nonlinear}. Following that, black holes with nonlinear electrodynamics have been widely constructed (See Refs \cite{kruglov2017nonlinear,amirabi2020black,pantig2022shadow,kruglov2022ads,gullu2021black,flores2021black}).
	
	One of the problems of black holes with electrical charge, and black holes with NED having at weak-field limit Maxwell's electrodynamics, is singularity at the center of the black hole \cite{amirabi2020black}. To solve this singularity problem, several models have been constructed using a magnetic charge distribution, and they are called regular black holes. Moreover, Bronnikov \cite{bronnikov2001regular} showed that regular electric black hole solution doesn't exist in the gravity coupled with a nonlinear electrodynamics which yields Maxwell's theory in the weak-field limit. One of the regular magnetic black hole, the Bardeen black hole, has an event horizon which satisfies the weak energy condition\cite{bardeen1968non,Mahamat2018}. The Bardeen solution has been reobtained by introducing an energy-momentum tensor, considered as the gravitational field of some sort of a nonlinear magnetic monopole charge $Q$\cite{ayon2000bardeen}. This kind of solution can also be called nonlinear magnetic-charged black hole.  Hereby, this is why this alternative has received considerable attention \cite{breton2005stability,abdujabbarov2013charged,ruffini2013einstein,lim2015motion,allahyari2020magnetically}.

%Black holes represent one of the most fascinating objects studied in astrophysics and cosmology. The mathematical framework necessary to study them is the General theory of relativity, established by Einstein\cite{einstein1916The}. At the center of the spacetime deformed by a black hole there is a singularity, at which both curvature and density becomes infinite, and physical laws are broken down\cite{hawking1973large}. Also, it is believed that a spacetime with singularity would appear with a well constructed quantum gravity theory. To solve this singularity problem, several models have been constructed, and they are called regular black holes. One of them, the Bardeen black hole, has an event horizon which satisfies the weak energy condition\cite{bardeen1968non,Mahamat2018}. The Bardeen solution has been reobtained by introducing an energy-momentum tensor, considered as the gravitational field of some sort of a nonlinear magnetic monopole charge $Q$\cite{ayon2000bardeen}. This kind of solution can also be called nonlinear magnetic-charged black hole.  Hereby, this is why this alternative has received considerable attention \cite{breton2005stability,abdujabbarov2013charged,ruffini2013einstein,lim2015motion,allahyari2020magnetically}.

Since the establishment of the black holes mechanical laws and their analogy between the thermodynamic laws, it has been suggested that black holes can be studied as  thermodynamic objects, having a temperature and an entropy, and moreover a volume, a heat capacity, and so on\cite{bekenstein1975statistical,bekenstein1973extraction,bardeen1973four,hawking1975particle}. This is why the black hole thermodynamic has been topic of study in many works\cite{tharanath2013thermodynamics,gibbons1977cosmological,yi2011thermodynamic,ghaderi2016thermodynamics,shahjalal2019thermodynamics,Banerjee2011b,appels2016thermodynamics}. Another interesting feature of black holes when studying the thermodynamics is the phase transition. Indeed, it has been shown that black holes undergo a phase transition ; in the AdS/CFT correspondence, through the black hole heat capacity~\cite{tharanath2013thermodynamics}.  Hence, it is possible  to see how the black hole behaves after a phase transition (see~\cite{rodrigue2020thermodynamic,cai2009thermodynamics,Mahamat2018,rodrigue2018thermodynamics,tharanath2013thermodynamics,tharanath2014phase,li2020thermodynamic}). Furthermore, through the Ehrenfest classification, the black hole can undergo a  first or second-order phase transition, as a discontinuity appears on the plot of the fisrt or second derivative of the free enthalpy. For example, one of the second derivative of the free enthalpy is the heat capacity, which is necessary to study the thermodynamic stability of the black hole. Indeed if the heat capacity negative(or positive), then the black hole is unstable(or stable). 

Nowadays black holes are also considered as very small, especially those which have been formed right after the Big Bang, called primordial black holes\cite{khlopov2010primordial}. According to their sizes, it is necessary to take into account quantum theory of gravity. The logarithmic approach is one of the predicted model considered as a result of quantum corrections\cite{das2002general,Sudhanshu2022corrected,abbas2023thermal}. Indeed, it has been introduced to investigate what the leading-order corrections are, when the size of the black hole is reduced. Therefore, this correction has been widely studied for many black holes. For example, Upadhyay et \textit{al.} studied the effect of correction parameter on thermodynamic behavior of a static black hole in $f(R)$ gravity\cite{upadhyay2021perturbed}. Other studies have been done taking into account thermal fluctuation in charged rotating black holes\cite{upadhyay2018leading}, regular black holes\cite{singh2021logarthmic} and Horava-Lifshitz black holes\cite{pourhassan2018quantum}. This motivates us to study how quantum corrections can affect the thermodynamic behavior of the nonlinear magnetic-charged black hole.

According to the standard model, the Universe is filled with a strange matter called dark matter, which constitutes about 23$\%$ of the total mass-energy of the universe\cite{xu2018kerr}. Its effects are present on galaxies, where it makes the outer parts of galaxies rotate faster than expected from their starlight and from the theoretical predictions Einstein general relativity. As theoretical candidates of dark matter, we have Cold Dark Matter (CDM)\cite{navarro1997universal}, Warm Dark Matter~\cite{dutta2021decaying,ruiz2021scalar} and Scalar Field Dark Matter~\cite{paranjape2021quantum,padilla2021consequences}. Another solution, the perfect fluid dark matter is also widely used, because it has been shown that perfect fluid dark matter can explain the asymptotically flat rotation curves concerning spiral galaxies\cite{siddhartha2003quintessence}. This has encouraged many works to consider the perfect fluid dark matter on the study of black holes\cite{saurabh2021imprints,shaymatov2021testing,ghosh2021charged,ma2021shadow}.

%In \cite{nam2018on}, Nam studied the effects of quintessence dark energy on the nonlinear magnetic charged black hole. Furthermore, in our recent work(see \cite{ndongmo2021thermodynamic}), we studied the effects of both quintessence dark energy and perfect fluid dark matter on this black hole. However, what could we have if we take into account only the perfect fluid dark matter?

In this paper, we put out the effects of perfect fluid dark matter and thermal fluctuation on the thermodynamic behavior of the nonlinear magnetic-charged black hole. 

The paper is organized as follows. First, through the horizon propriety and the surface gravity definition, we will determine the black hole mass and  Hawking temperature of the black hole. Secondly, we will use them in order to find the corrected entropy due to quantum fluctuations, and then analyze the effects of perfect fluid dark matter and corrected parameter. Then, we will analyze the thermodynamic stability of the black hole through the evolution of the specific heat and Gibbs free energy, then we see which role plays the corrected parameter. We will end by a conclusion.

\section{Hawking Temperature, corrected entropy and Gibbs free energy}

The metric of static spherically symmetric solution for the Einstein equations describing the nonlinear magnetic-charged black hole in the background of perfect fluid dark matter is expressed as \cite{ma2021shadow,ndongmo2023thermodynamics}

\begin{equation}\label{}
ds^2=-f(r)dt^2+\frac{1}{f(r)}dr^2+r^2(d\theta^2+\sin^2\theta d\phi^2),
\end{equation}
with  $f(r)=1-\frac{2Mr^2}{r^3+Q^3}+\frac{\alpha}{r}\ln\frac{r}{|\alpha|}$.

Here, $M$ the black hole mass, $Q$ the magnetic charge and $\alpha$ is the dark matter parameter. Using the horizon propriety~\cite{nam2018on,tharanath2014phase}, and solving the following equation at the horizon

\begin{equation}
f(r_h)=0,												
\end{equation}
leads to
\begin{equation}\label{Ma2}
M=\frac{(r_h^3+Q^3)}{2r_h^2}\left(1+\frac{\alpha}{r_h}\ln\frac{r_h}{|\alpha|}\right).
\end{equation}

Eq. (\ref{Ma2}) gives the relation between the black hole mass and its horizon radius. Here, we will focus on the event horizon to make the thermodynamic analysis. 

%These intensive thermodynamic variables are respectively defined as
%
%\begin{equation}\label{intensive_var2}
%T_h=\left(\frac{\partial M}{\partial S}\right)_{Q,\alpha}, \ \Phi_h=\left(\frac{\partial M}{\partial Q}\right)_{S,\alpha}, \ \beta_h=\left(\frac{\partial M}{\partial \alpha}\right)_{S,Q}.
%\end{equation}

%In the first expression of \eqref{intensive_var2}, we can find the formula of the entropy, expressed as follows
%\begin{equation}\label{entro3}
%S=\int\frac{1}{T_h}dM=\int\frac{1}{T_h}\frac{\partial M}{\partial r_h}dr_h.
%\end{equation} 

%Now, we are going to compute, $T_h$ and $\frac{\partial M}{\partial r_h}$, which will help us to find the entropy.

Through the surface gravity definition at the horizon\cite{nam2018on}, the Hawking temperature $T_h$ is given by
\begin{equation}
T_h= \frac{\kappa}{2\pi}=\frac{f'(r_h)}{4\pi}.
\end{equation}

Taking into account Eq. \eqref{Ma2} and the expression of the metric function $f(r)$, we obtain the following Hawking temperature
%\begin{widetext}
\begin{equation}\label{Temp_nonlin2}
T_h=\frac{1}{4\pi (r_h^3+Q^3)}\left[\frac{r_h^3-2Q^3}{r_h}+\frac{\alpha}{r_h^2}\left( r_h^3+Q^3-3Q^3\ln\left(\frac{r_h}{|\alpha|}\right) \right)  \right].
\end{equation}
\begin{figure}[h]
	
	\centering
	\includegraphics[scale=0.45]{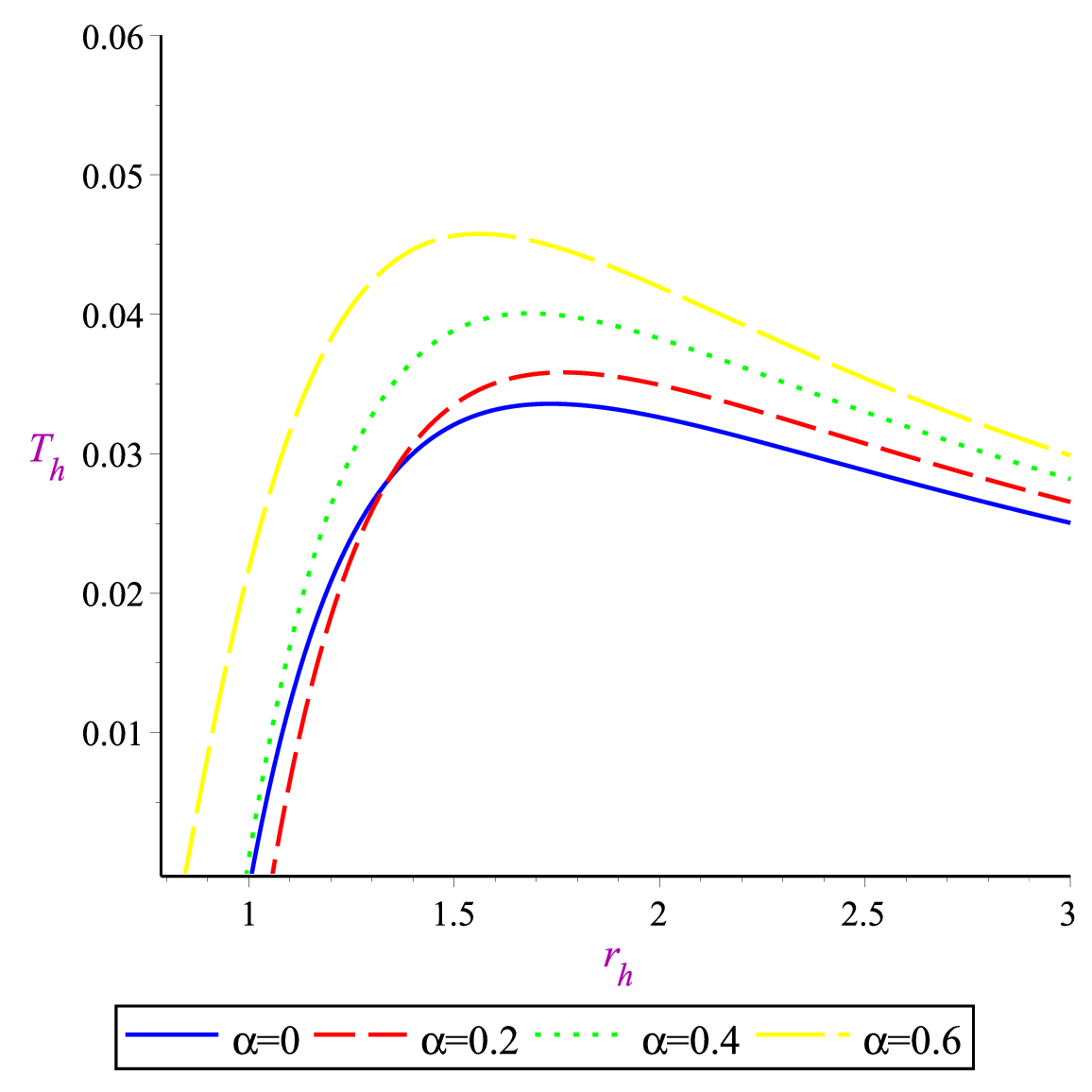}
	\caption{\label{temperature_curve_DM}Change of the black hole temperature $T_h$ for different values of dark matter parameter with $Q=0.8$.}
\end{figure}
%	\end{center}
%\end{widetext}

%\begin{widetext}
%	\begin{center}
%		\begin{figure}[!h]
%			\begin{minipage}[!h]{6cm}
%				\centering
%				\includegraphics[scale=0.35]{C_alpha2.eps}
%				(a) for $Q=0$
%				%	\caption{Variation of the black hole temperature $T$ in the presence of
%				%		quintessence dark energy with characteristics $(Q,c,\epsilon)=(0,0.02,-0.5)$. Here $a=0.3$ corresponds to blue continuous line, $a=0.6$ to red dash and $a=0.9$ to green dash-point.}
%			\end{minipage}
%			\hspace{3cm}
%			\begin{minipage}[!h]{6cm}
%				\centering
%				\includegraphics[scale=0.35]{C_alpha1.eps}
%				(b) for $Q=1$
%				%	\caption{Variation of the black hole temperature $T$ in the presence of
%				%		quintessence dark energy with characteristics $(Q,c,\epsilon)=(1,0.02,-0.5)$. Here $a=0.3$ corresponds to blue continuous line, $a=0.6$ to red dash and $a=0.9$ to green dash-point.}
%			\end{minipage}
%			\caption{\label{heat_capacity_alpha}Change of the heat capacity $C$ of the nonlinear magnetic-charged black hole in the background of perfect fluid dark matter.}
%		\end{figure}
%	\end{center}	
%\end{widetext}	

In figure \eqref{temperature_curve_DM}, we plot the temperature of the nonlinear magnetic-charged black hole surrounded by perfect fluid dark matter. On this plot, we can see that the black hole temperature increases and reaches a maximum, before having a phase of decrease.
%, which is similar to the behaviour of the temperature of the nonlinear magnetic-charged black hole surrounded by quintessence, plotted in \eqref{Temp_nonlin_plot}.
Furthermore, this figure shows us that this maximum increases for higher values of dark matter parameter $\alpha$. Let us notice that the case $\alpha=0$ corresponds to the temperature of black hole without perfect fluid dark matter.

%In figure \eqref{temperature_curve_DM}, we plotted the temperature of the nonlinear magnetic-charged black hole surrounded by perfect fluid dark matter. On that plot, looking at subfigure \eqref{temperature_curve_DM}(b), we can see that the black hole temperature increases and reaches a maximum, and then decreases.
%%, which is similar to the behaviour of the temperature of the nonlinear magnetic-charged black hole surrounded by quintessence, plotted in \eqref{Temp_nonlin_plot}.
% Furthermore, this maximum increases for higher values of dark matter parameter. 
%
%However, looking at subfigure \eqref{temperature_curve_DM}(a), we can see that the temperature exists for lower values of horizon radius, but this behaviour does not appear in the case of absence of perfect fluid dark matter(See figure \eqref{temperature_curve_DM}(c)). Therefore, this result means that dark matter could lead small black holes to exist by having temperature.

%In the other hand, comparing figure \eqref{Temp_nonlin_plot} and subfigure \eqref{temperature_curve_DM}(b), we notice that the maximum of the temperature shifts to lower values as we increase the quintessence parameter $c_q$ and decrease the perfect fluid dark matter parameter $\alpha$.

%\section{III. The corrected entropy and effects of the corrected parameter}
From the black hole studied here, we need to write the first law of the black hole thermodynamics and then find out the entropy before computing the corrected entropy. The first law is expressed as\cite{nam2018on}
\begin{equation}\label{first_law3}
dM = T_{h}dS_0+\Phi_{h} dQ + \beta_{h}d\alpha,
\end{equation}
where $S_0$ represents the entropy at the equilibrium without considering thermal fluctuation, or the uncorrected entropy. Let us notice that $S_0$, the magnetic charge $Q$ and the dark matter parameter $\alpha$ form a complete set of extensible variables. $T_h$ is the Hawking temperature at the horizon and $\Phi_h$ is the potential. $\beta_h$ is the conjugating quantity of the dark matter parameter $\alpha$. 

Now, let us compute the uncorrected entropy of the black hole. To having it, we can write first the differential of $M(S_0,Q,\alpha)$ as

\begin{equation}\label{derivMa1}
dM(S_0,Q,\alpha)=\left.\frac{\partial M}{\partial S_0}\right|_{Q,\alpha}dS_0+\left.\frac{\partial M}{\partial Q}\right|_{S_0,\alpha}dQ+\left.\frac{\partial M}{\partial \alpha}\right|_{S_0,Q}d\alpha.
\end{equation}

Proceeding by identification between \eqref{derivMa1} and \eqref{first_law3}, we have
\begin{equation}\label{ident1}
\Phi_h=\left.\frac{\partial M}{\partial Q}\right|_{S_0,\alpha} \ \text{and} \ \beta_h=\left.\frac{\partial M}{\partial \alpha}\right|_{S_0,Q}.
\end{equation}

However, the mass $M(r_h,Q,\alpha)$ expressed in \eqref{Ma2} can be differentiated as

\begin{equation}\label{derivMa2}
dM(r_h,Q,\alpha)=\left.\frac{\partial M}{\partial r_h}\right|_{Q,\alpha}dr_h+\left.\frac{\partial M}{\partial Q}\right|_{r_h,\alpha}dQ+\left.\frac{\partial M}{\partial \alpha}\right|_{r_h,Q}d\alpha.
\end{equation}

Therefore, since \eqref{derivMa2} and \eqref{first_law3} are equal, we straightforwardly obtain the following equation
\begin{equation}\label{22}
\begin{array}{r c l}
&&T_hdS_0-\left.\frac{\partial M}{\partial r_h}\right|_{Q,\alpha}dr_h+\left(\Phi_h-\left.\frac{\partial M}{\partial Q}\right|_{r_h,\alpha}\right)dQ\\
&&+\left(\beta_h-\left.\frac{\partial M}{\partial \alpha}\right|_{r_h,Q}\right)d\alpha=0.
\end{array}
\end{equation}

However, taking into account \eqref{ident1} leads us to have
\begin{equation}
T_hdS_0=\frac{\partial M}{\partial r_h}dr_h.
\end{equation}

Finally, the entropy can be computed through the following equation\cite{nam2018on,nam2018non}
\begin{equation}\label{entro}
S_0=\int\frac{1}{T_h}\frac{\partial M}{\partial r_h}dr_h.
\end{equation}

To compute this, we have to find $\frac{\partial M}{\partial r_h}$ from Eq. \eqref{Ma2}, on that way

%\begin{widetext}
\begin{eqnarray*}
	\begin{array}{r c l}
		\frac{\partial M}{\partial r_h}&=&\frac{1}{2}\left\{ \frac{3r_h^3-2(r_h^3+Q^3)}{r_h^3}\left(1+\ln\left(\frac{r_h}{|\alpha|}\right)\right)+\left(\frac{r_h^3+Q^3}{r_h^2}\right)\left(-\frac{\alpha}{r_h^2}\ln\left(\frac{r_h}{|\alpha|}\right)+\frac{\alpha}{r_h}\times\frac{\frac{1}{|\alpha|}}{\frac{r_h}{|\alpha|}}\right) \right\}\\
		&=&\frac{1}{2}\left\{3+3\frac{\alpha}{r_h}\ln\left(\frac{r_h}{|\alpha|}\right)-2-\frac{2Q^3}{r_h^3}-\frac{3\alpha(r_h^3+Q^3)}{r_h^4}\ln\left(\frac{r_h}{|\alpha|}\right)+\frac{\alpha(r_h^3+Q^3)}{r_h^4} \right\}\\
		&=&\frac{1}{2}\left\{ 1-\frac{2Q^3}{r_h^3}+\frac{\alpha}{r_h^2}\left[ \frac{r_h^3+Q^3}{r_h^2}-\frac{3Q^3}{r_h^2}\ln\left(\frac{r}{|\alpha|}\right) \right] \right\}.
	\end{array}
\end{eqnarray*}
%\end{widetext}

Hereby, the first derivative of the black hole mass is found as

\begin{equation}\label{deriv_mass3}
\frac{\partial M}{\partial r_h}=\frac{1}{2r_h^2}\left\{ \frac{r_h^3-2Q^3}{r_h}+\frac{\alpha}{r_h^2}\left[ r_h^3+Q^3-3Q^3\ln\left(\frac{r_h}{|\alpha|}\right) \right] \right\}.
\end{equation}
Introducing Eq.\eqref{deriv_mass3} and \eqref{Temp_nonlin2} into Eq. \eqref{entro}, we get the black hole entropy at the equilibrium expressed as
\begin{equation}\label{entro_b}
S_0=2\pi\int\left(\frac{r_h^3+Q^3}{r_h^2}\right)dr=\pi r_h^2\left(1-\frac{2Q^3}{r_h^3}\right).
\end{equation}

Here, we can notice that this result is the same than the one obtained in the presence of quintessence dark energy, found in by Nam\cite{nam2018on}, and also in our recent work where we putted together quintessence and perfect fluid dark matter\cite{ndongmo2023thermodynamics}. Hereby, we can say that neither perfect fluid dark matter does not affect the evolution of entropy of nonlinear magnetic charged black hole.

Now, we will compute the expression of corrected entropy at the equilibrium $S$, using the formula found in Ref \cite{Sudhanshu2022corrected}

\begin{equation}\label{corrected_S}
S=S_0-\beta\ln(S_0T_h).
\end{equation}
Here, $\beta$ is called the corrected parameter, and has only two values. If $\beta=0$, Eq. \eqref{corrected_S} describes the uncorrected entropy, and for $\beta=\frac{1}{2}$, Eq. \eqref{corrected_S} describes the corrected entropy due thermal fluctuation. Thus, we obtain

%\begin{widetext}
\begin{equation}\label{corrected_S2}
S=\pi r_h^2\left(1-\frac{2Q^3}{r_h^3}\right)-\beta\ln\left\{  \frac{\left	(1-\frac{2Q^3}{r_h^3}\right)}{16\pi (r_h^3+Q^3)^2}\left[\frac{r_h^3-2Q^3}{r_h}+\frac{\alpha}{r_h^2}\left( r_h^3+Q^3-3Q^3\ln\left(\frac{r_h}{|\alpha|}\right) \right)  \right]^2 \right\}.
\end{equation}
%\end{widetext}

In order to have a better appreciation of the impact of thermal fluctuation on the black hole entropy, we plotted it in figure (2). Analyzing this plot, we remark that for higher values of the horizon radius, we have a linear increase of entropy ; as well as if the system is in equilibrium. However, at lower values of the horizon radius, while the equilibrium entropy or the entropy without correction($\beta=0$) increases, the corrected entropy($\beta=0.5, $) shows a phase of decrease with messy behavior of the entropy for smaller values of the dark matter parameter. This result means that the thermal fluctuations violates the second law of thermodynamics, as it is also showed in \cite{Sudhanshu2022corrected,abbas2023thermal,upadhyay2018leading,pourhassan2019thermodynamics,ganai2019thermal}. Furthermore, we notice that the effect of thermal fluctuation can be neglected for larger black holes.

%\begin{center}
\begin{figure}[h]
	\centering
	\includegraphics[scale=0.50]{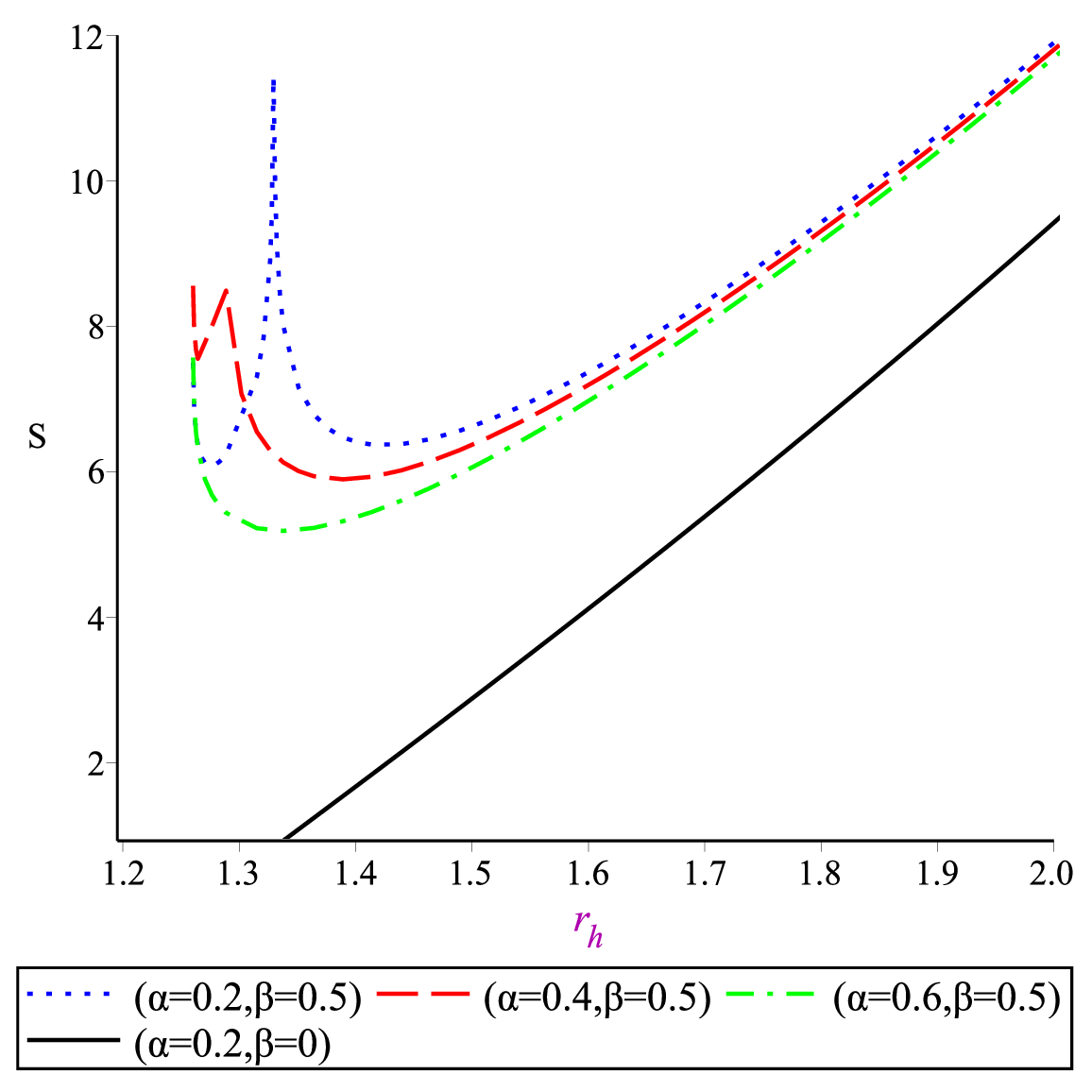}
	\caption{\label{Phi_2}Variation of the corrected  entropy $S$ for different values of  $\alpha$ and $\beta$.} 
\end{figure}

\section{Thermodynamic stability}

Here, the thermodynamic stability of the black hole will be studied through  the calculus and plot of the corresponding heat capacity including thermal fluctuation, which is found through the formula

\begin{equation}
C=T_h\left(\frac{\partial S}{\partial T_h}\right)_{Q,\alpha}=T_h\left(\frac{\partial S}{\partial r_h}\frac{\partial r_h}{\partial T_h}\right)_{Q,\alpha}.
\end{equation}
After computing it, we get the following expression
%\begin{widetext}
\begin{eqnarray}
\begin{array}{r c l}
C&=&\frac{A}{B},\\
\text{with} \ A&=& 2\left(r_h^4-2r_hQ^3+\alpha r_h^3+\alpha Q^3\left(1-\ln\left(\frac{r_h}{|\alpha|}\right)\right)\right)((6\pi Q^{12}+9\pi Q^9r_h^3-3\pi Q^3r_h^9+15Q^9\beta r_h\\
&+&30Q^6\beta r_h^4-12Q^3\beta r^7)\alpha\ln\left(\frac{r_h}{|\alpha|}\right)-2\pi Q^{12}\alpha+4\pi Q^{12} r_h-5\pi Q^9\alpha r_h^3+4\pi Q^9 r_h^4-3\pi Q^6\alpha r^7\\
&+&\pi Q^3\alpha r_h^9-2\pi Q^3r_h^{10}+\pi\alpha r_h^{12}+\pi r_h^{13}-11Q^9\alpha \beta r_h+6Q^9\beta r_h^2-12Q^6\alpha \beta r_h^4+21Q^6\beta r_h^5\\
&-&12Q^3\beta r_h^8+\alpha \beta r_h^{10}, \\
\text{and} \ B&=&   r_h(2Q^3-r_h^3)(3Q^3\alpha\ln\left(\frac{r_h}{|\alpha|}\right)-\alpha Q^3+2Q^3r_h-\alpha r_h^3-r_h^4)(6Q^6\alpha\ln\left(\frac{r_h}{|\alpha|}\right)\\
&+&15Q^3r_h^3\alpha\ln\left(\frac{r_h}{|\alpha|}\right)-5Q^6\alpha+2Q^6r_h-7Q^3\alpha r_h^3+10Q^3r_h^4-2\alpha r_h^6-r_h^7).
\end{array}
\end{eqnarray}	
%\begin{equation}
%C=-\frac{2\, \left( 3\,\ln  \left(  \left| \alpha \right|  \right) {Q}^{3}
%	\alpha-3\,\ln  \left( r \right) {Q}^{3}\alpha+{Q}^{3}\alpha-2\,{Q}^{3}
%	r+\alpha\,{r}^{3}+{r}^{4} \right) \pi \, \left( {Q}^{3}+{r}^{3}
%	\right) ^{2}
%}{A},
%\end{equation}
%where 
%
%\[ A=r ( 6\,{Q}^{6}\alpha\,\ln  \left(  \left| \alpha \right| 	\right) +15\,{Q}^{3}\alpha\,{r}^{3}\ln  \left(  \left| \alpha	\right|  \right) -6\,{Q}^{6}\alpha\,\ln  \left( r \right) -15\,{Q}^{3}\alpha\,{r}^{3}\ln  \left( r \right) \]		
%\[+5\,{Q}^{6}\alpha-2\,{Q}^{6}r+7\,{Q}^{3}\alpha\,{r}^{3}-10\,{Q}^{3}{r}^{4}+2\,\alpha\,{r}^{6}+{r}^{7}). \]
%\end{widetext}

\begin{center}
	\begin{figure}[h]
		\hspace{-1cm}
		\begin{minipage}[h]{4cm}
			\centering
			\includegraphics[scale=0.35]{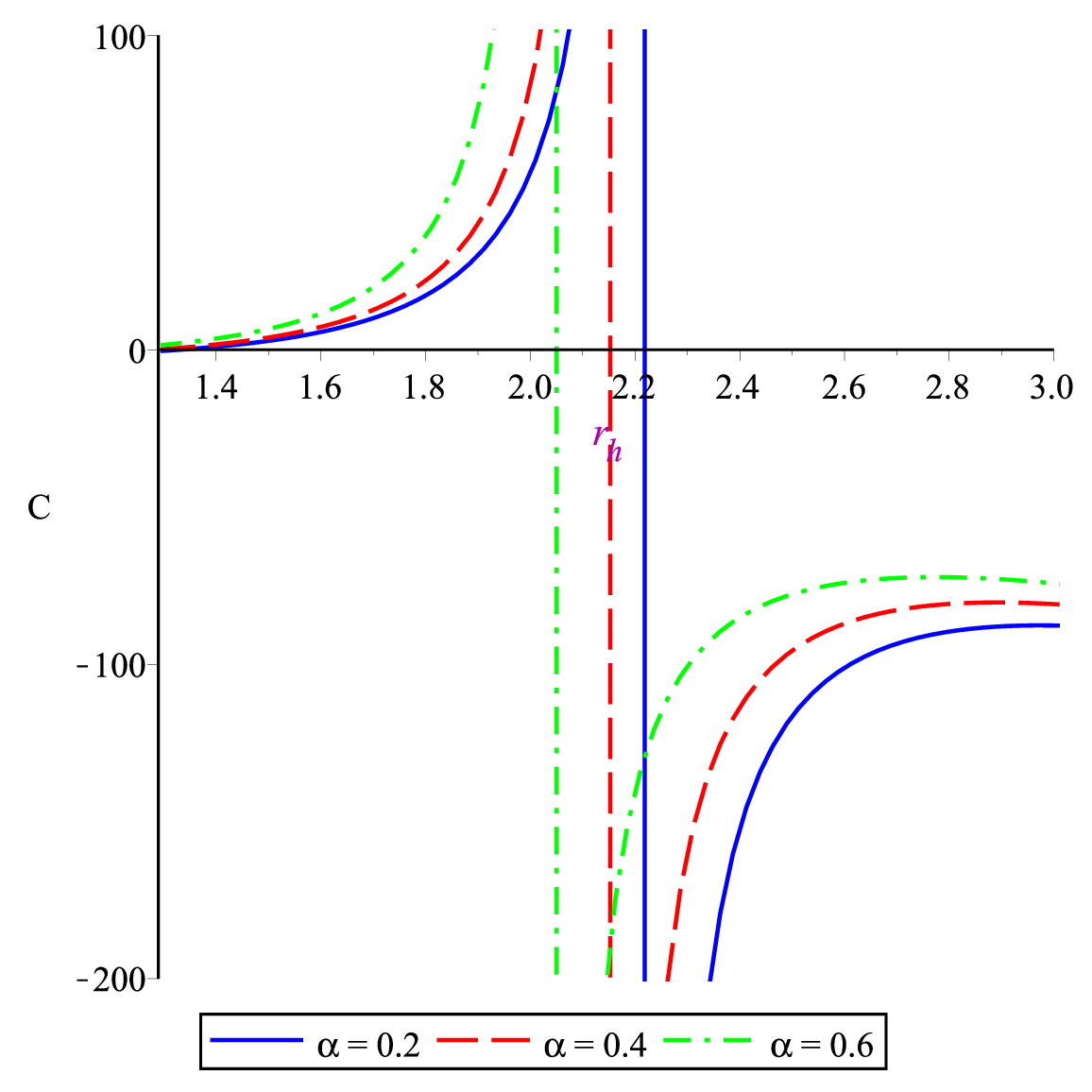}
			(a) for $\beta=0$
			%	\caption{Variation of the black hole temperature $T$ in the presence of
			%		quintessence dark energy with characteristics $(Q,c,\epsilon)=(0,0.02,-0.5)$. Here $a=0.3$ corresponds to blue continuous line, $a=0.6$ to red dash and $a=0.9$ to green dash-point.}
		\end{minipage}
		\hspace{3cm}
		\begin{minipage}[h]{3cm}
			\centering
			\includegraphics[scale=0.35]{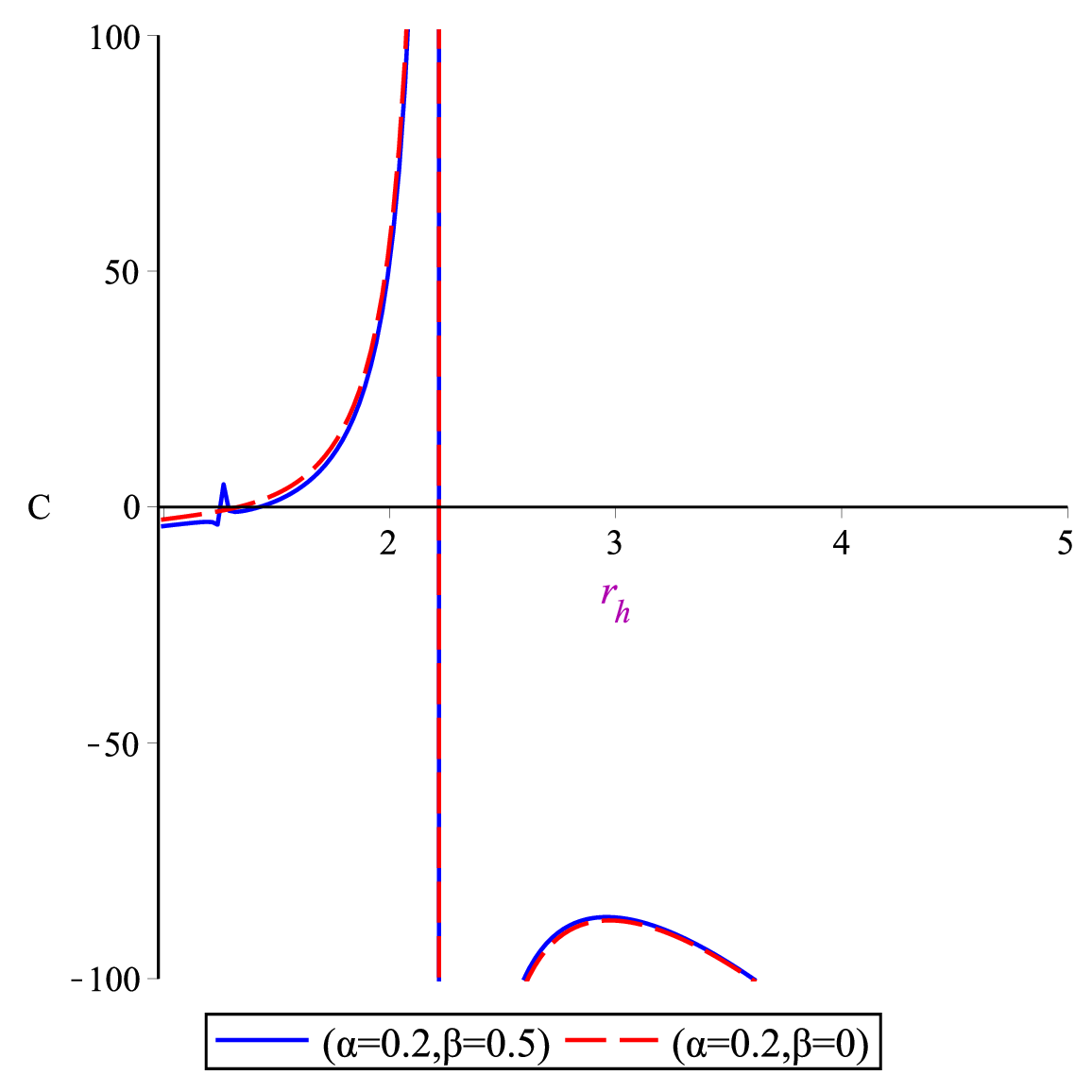}
			(b) 
			%	\caption{Variation of the black hole temperature $T$ in the presence of
			%		quintessence dark energy with characteristics $(Q,c,\epsilon)=(1,0.02,-0.5)$. Here $a=0.3$ corresponds to blue continuous line, $a=0.6$ to red dash and $a=0.9$ to green dash-point.}
		\end{minipage}
		
		\hspace{3cm}
		\begin{minipage}[h]{3cm}
			\centering
			\includegraphics[scale=0.35]{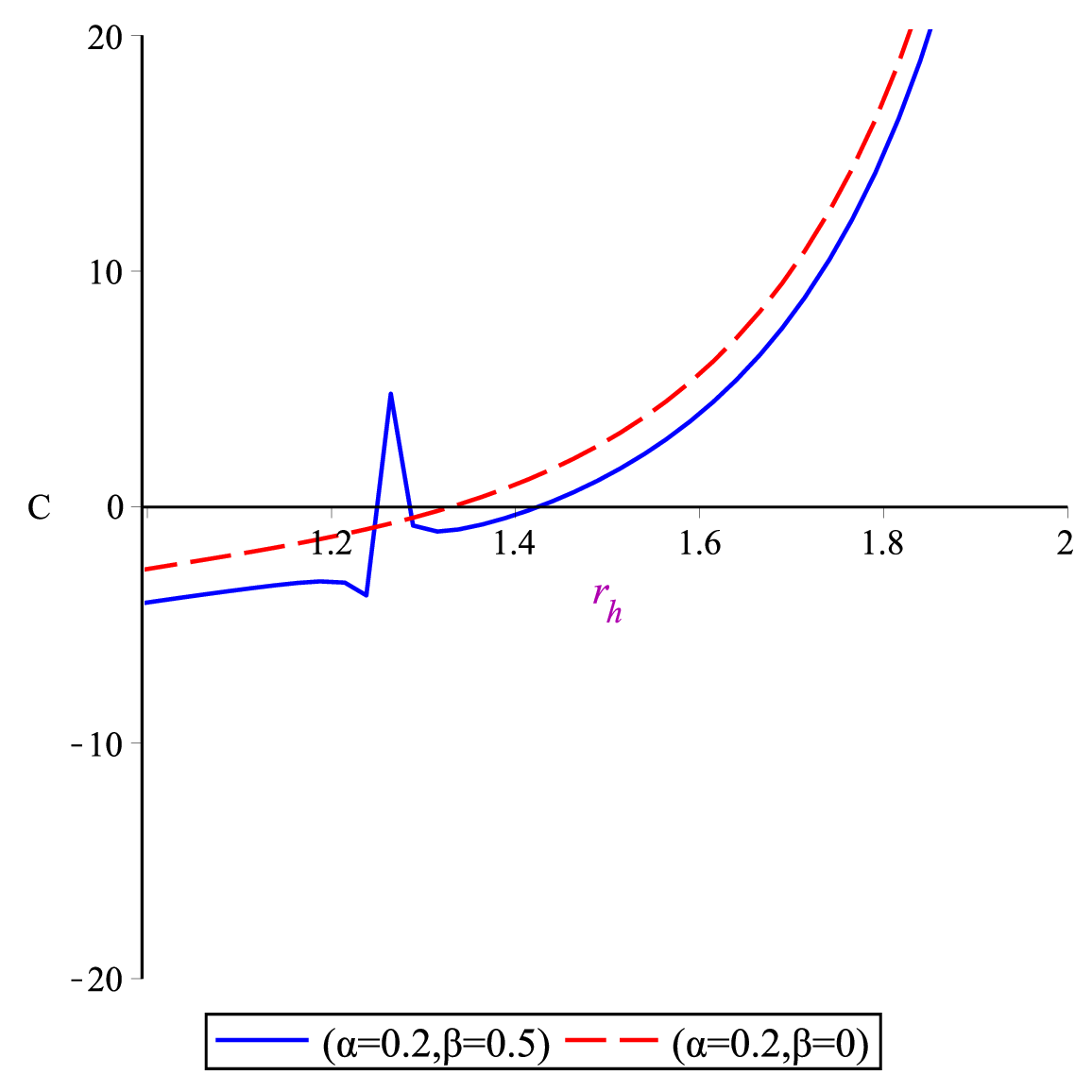}
			(c)
			%	\caption{Variation of the black hole temperature $T$ in the presence of
			%		quintessence dark energy with characteristics $(Q,c,\epsilon)=(1,0.02,-0.5)$. Here $a=0.3$ corresponds to blue continuous line, $a=0.6$ to red dash and $a=0.9$ to green dash-point.}
		\end{minipage}
		\hspace{4cm}
		\caption{\label{heat_capacity_alpha}Change of the heat capacity $C$ of the nonlinear magnetic-charged black hole in the background of perfect fluid dark matter with and without quantum corrections for $Q=1$.}
	\end{figure}
\end{center}

In figure \eqref{heat_capacity_alpha}, we plot the heat capacity $C$ of the nonlinear magnetic-charged black hole in the background of perfect fluid dark matter. Analysing its plot, especially subfigure \eqref{heat_capacity_alpha}(a) which corresponds to the absence of quantum corrections, we can see the presence of a discontinuity for each dark matter parameter $\alpha$. Physically meaning, that is to say that the black hole undergoes a second-order phase transition. Then, this phase transition leads the black hole to move from the stable phase($C>0$) to the unstable phase($C<0$). Moreover, through subfigure \eqref{heat_capacity_alpha}(b), we see that thermal fluctuation does not modify the region of occurrence of the second-order phase transition. However, taking into account quantum corrections, then looking at subfigures \eqref{heat_capacity_alpha}(b) and (c),  we see that the heat capacity is not only an increasing function, but evolves with two picks. This implies that thermal fluctuation affects a smaller black hole.

\begin{center}
	\begin{figure}[h]
		\hspace{0cm}
		\begin{minipage}[h]{4cm}
			\centering
			\includegraphics[scale=0.32]{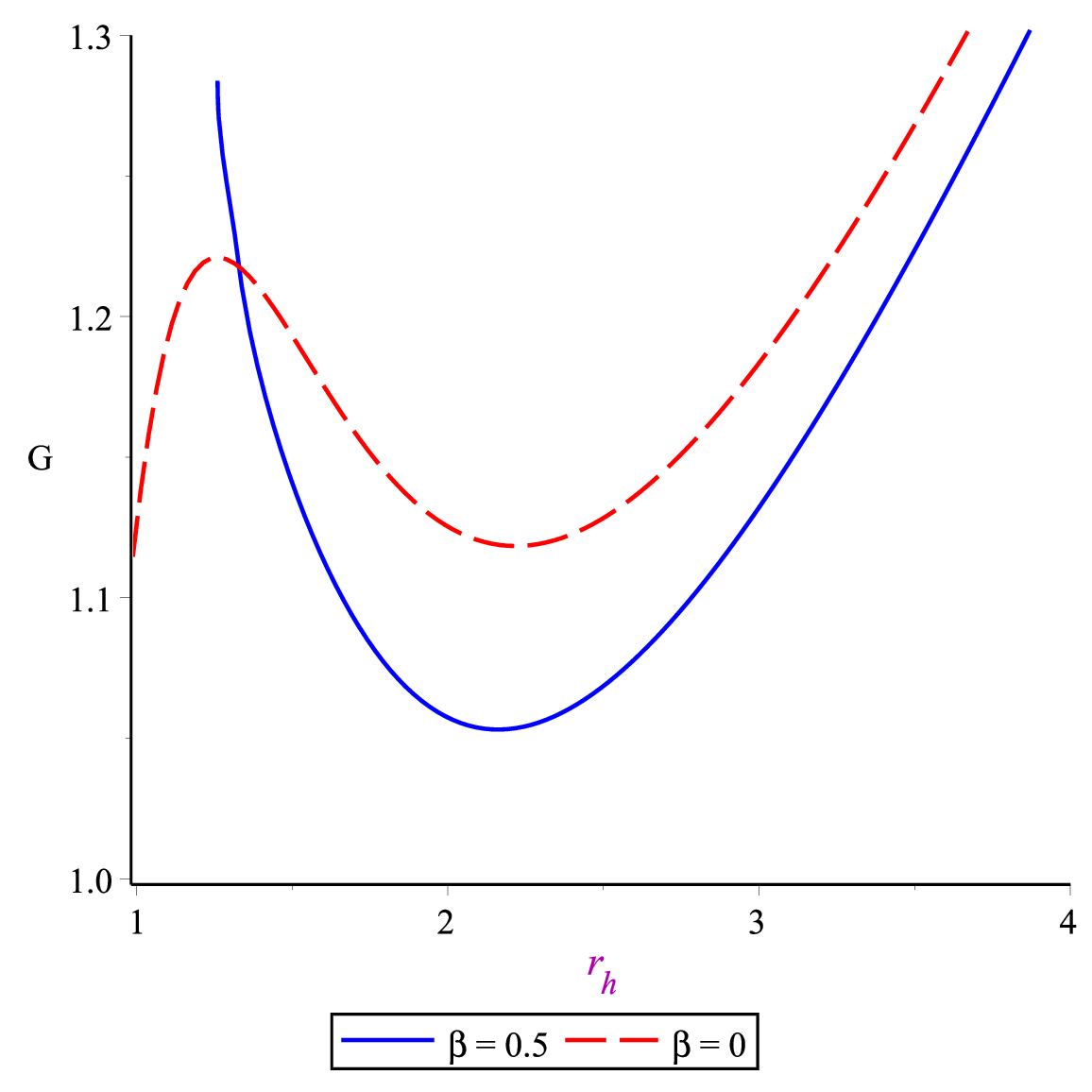}
			(a) for $\alpha=0.2$
			%	\caption{Variation of the black hole temperature $T$ in the presence of
			%		quintessence dark energy with characteristics $(Q,c,\epsilon)=(0,0.02,-0.5)$. Here $a=0.3$ corresponds to blue continuous line, $a=0.6$ to red dash and $a=0.9$ to green dash-point.}
		\end{minipage}
		\hspace{3cm}
		\begin{minipage}[h]{3cm}
			\centering
			\includegraphics[scale=0.32]{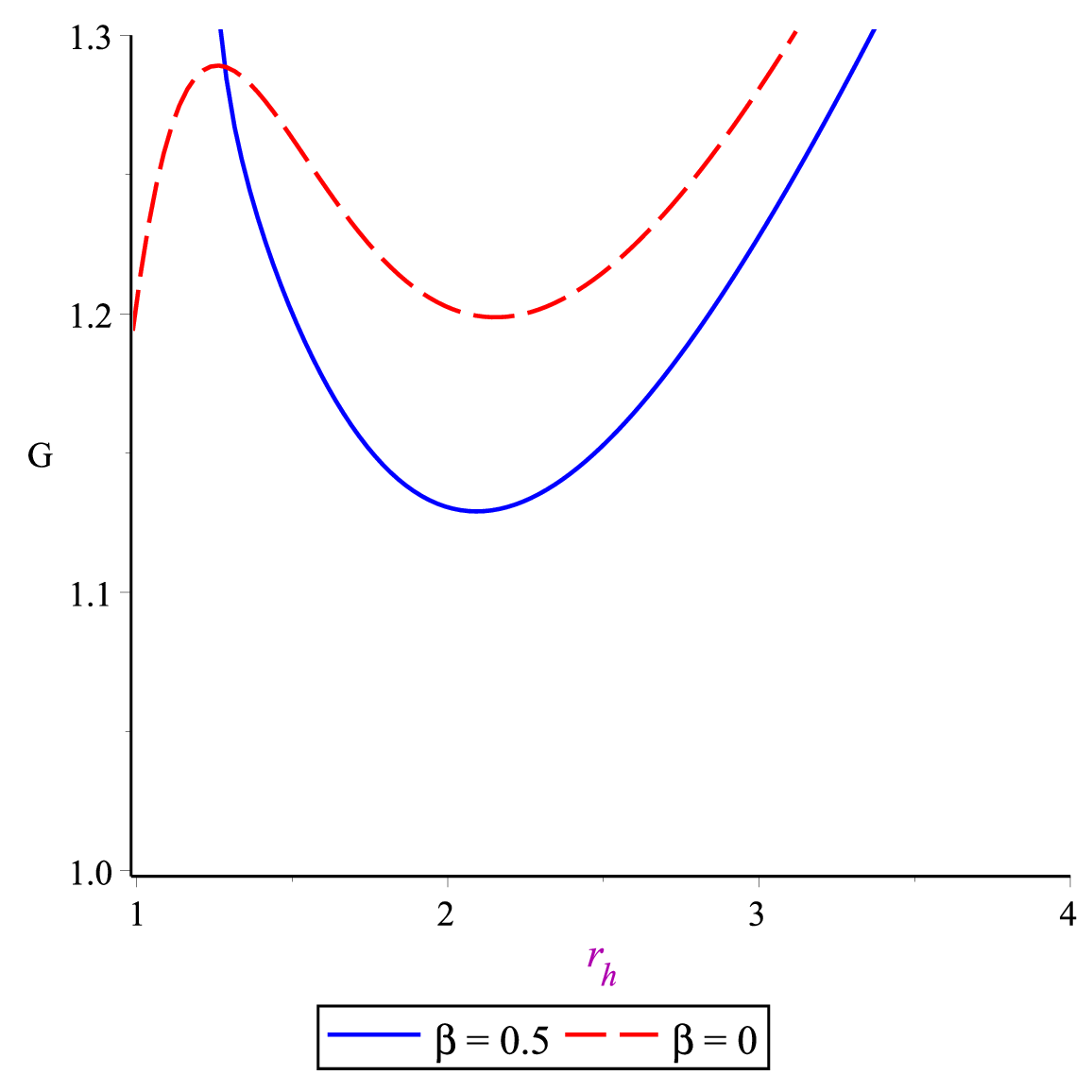}
			(b) for $\alpha=0.4$
			%	\caption{Variation of the black hole temperature $T$ in the presence of
			%		quintessence dark energy with characteristics $(Q,c,\epsilon)=(1,0.02,-0.5)$. Here $a=0.3$ corresponds to blue continuous line, $a=0.6$ to red dash and $a=0.9$ to green dash-point.}
		\end{minipage}
		%			\hspace{2cm}
		\caption{\label{Gibbs_plot}Variation of the Gibbs free energy $G$ of the nonlinear magnetic-charged black hole in the background of perfect fluid dark matter with and without quantum corrections for $Q=1$.}
	\end{figure}
\end{center}

Now, in order to better appreciate the effects of thermal fluctuation, we also depicted the Gibbs free energy $G$ in figure \eqref{Gibbs_plot}, which is expressed as follows 

\begin{equation}\label{Gibbs}
G=M-TS,
\end{equation}
where $M$ is the black hole mass given in Eq. \eqref{Ma2}, $T$ is the temperature given in Eq. \eqref{Temp_nonlin2} and the entropy $S$ is found in Eq. \eqref{corrected_S2}. Therefore, figure \eqref{Gibbs_plot}, shows that in the range $[1,4]$ of the horizon radius $r_h$, the Gibbs free energy has two extrema in the absence of thermal fluctuation (continuous blue line), while having only one extremum in the presence of thermal fluctuation (discontinuous red line). This result shows that  thermal fluctuation decreases the number of phases of change of Gibbs free energy $G$, by modifying its behavior for small horizon radii values. Furthermore, figure \eqref{Gibbs_plot}(b) show that this behavior remains the same when the dark matter parameter $\alpha$ is changed.

\begin{center}
	\begin{figure}[h]
		\hspace{0cm}
		\begin{minipage}[h]{4cm}
			\centering
			\includegraphics[scale=0.33]{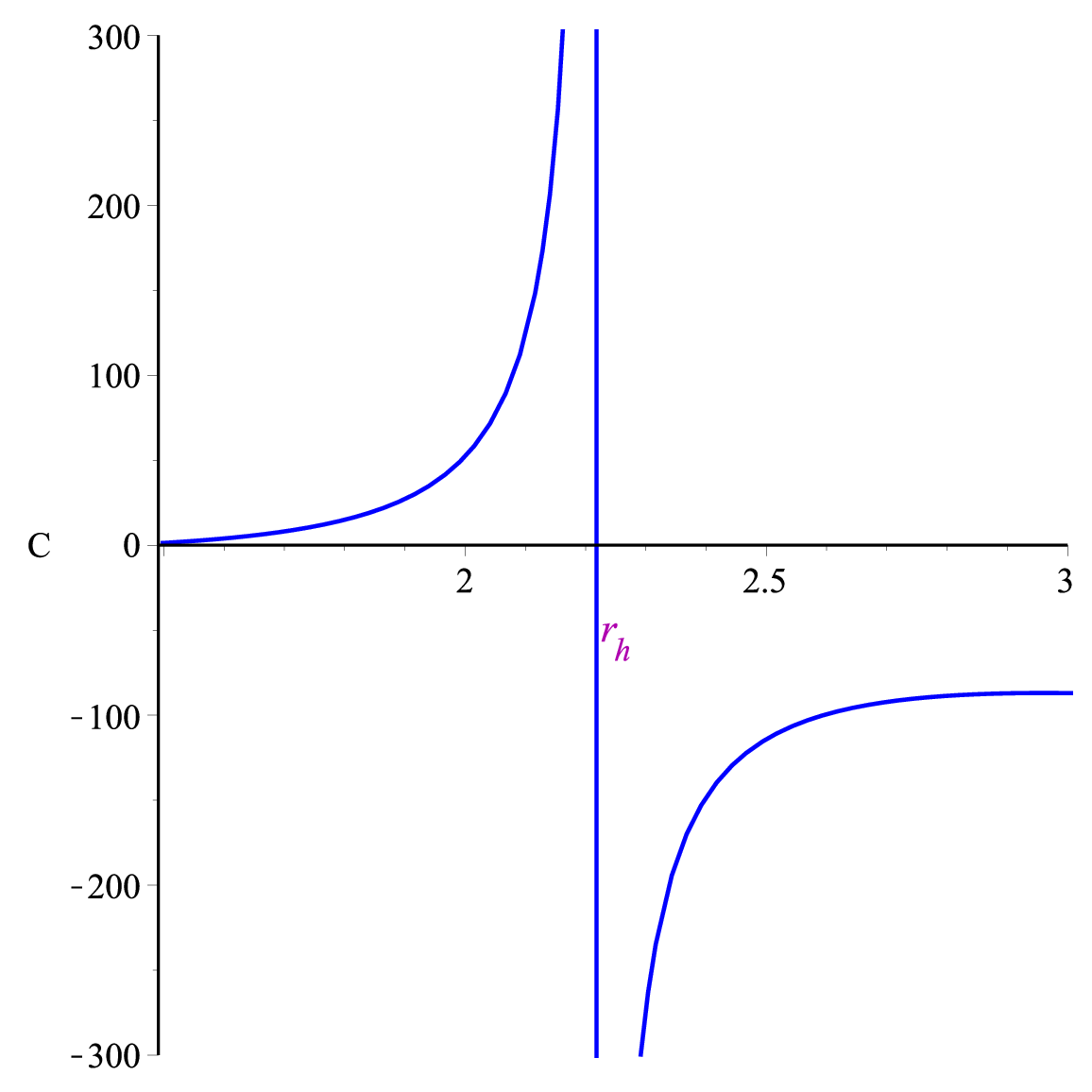}
			(a) Heat capacity $C$
			%	\caption{Variation of the black hole temperature $T$ in the presence of
			%		quintessence dark energy with characteristics $(Q,c,\epsilon)=(0,0.02,-0.5)$. Here $a=0.3$ corresponds to blue continuous line, $a=0.6$ to red dash and $a=0.9$ to green dash-point.}
		\end{minipage}
		\hspace{3cm}
		\begin{minipage}[h]{4cm}
			\centering
			\includegraphics[scale=0.33]{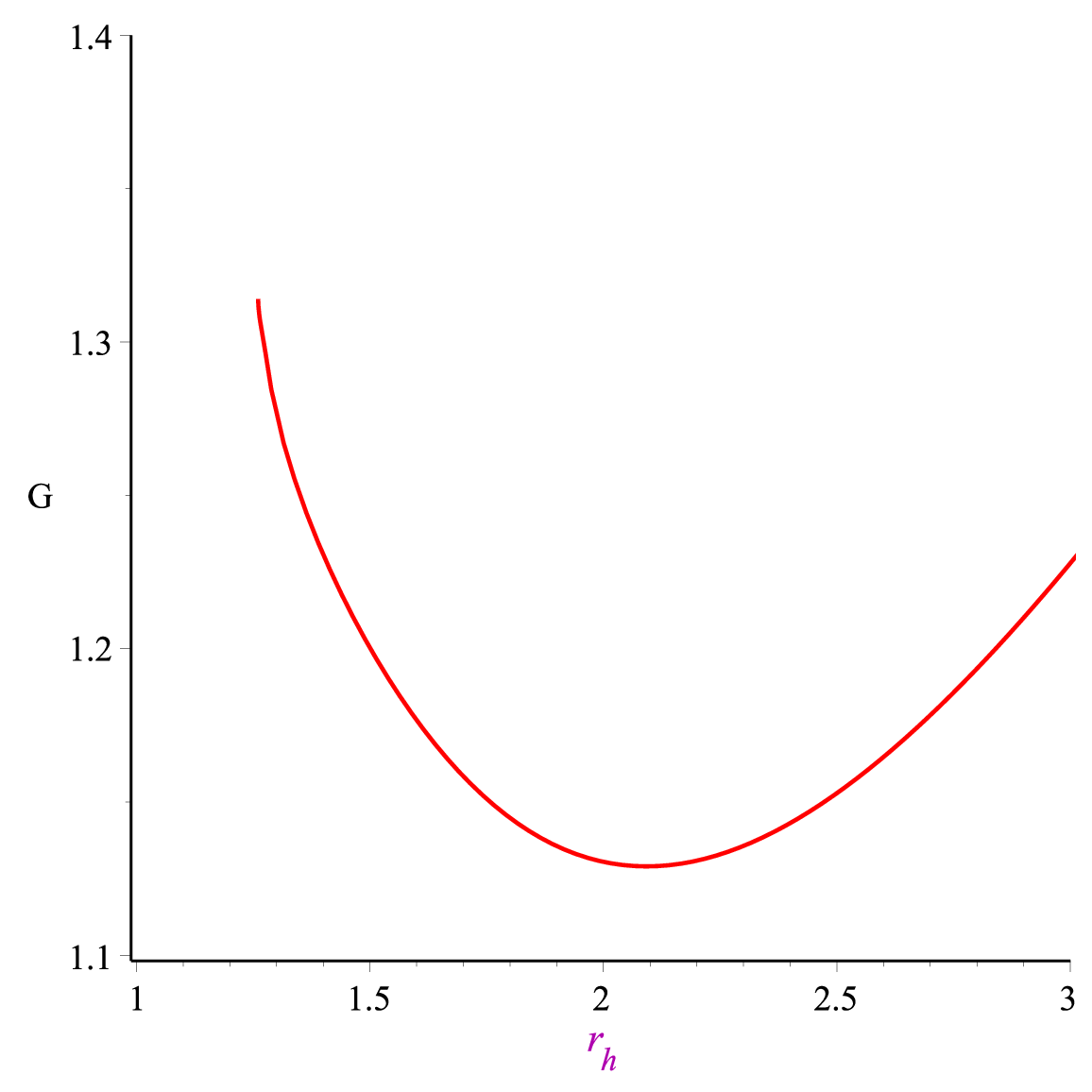}
			(b) Gibbs free energy $G$
			%	\caption{Variation of the black hole temperature $T$ in the presence of
			%		quintessence dark energy with characteristics $(Q,c,\epsilon)=(1,0.02,-0.5)$. Here $a=0.3$ corresponds to blue continuous line, $a=0.6$ to red dash and $a=0.9$ to green dash-point.}
		\end{minipage}
		%			\hspace{2cm}
		\caption{\label{C&G}Variation of the heat capacity $C$ and the Gibbs free energy $G$  for $(Q,\alpha,\beta_{h})=(1,0.4,0.5)$.}
	\end{figure}
\end{center}

To make further analysis of the stability, we plotted in figure \eqref{C&G} the heat capacity and Gibbs free energy, with same values of $(Q,\alpha,\beta_{h})$. Comparing subfigure \eqref{C&G}(a), with subfigure \eqref{C&G}(b), we can see that when the heat capacity is negative, the Gibbs free energy decreases, and when the heat capacity is positive, we have a phase of increase of Gibbs free energy. Hence this result means that a stable black hole has a decreasing Gibbs free energy, and an unstable black hole has an increasing Gibbs free energy. Therefore, its behaviour is changed through the second order phase transition.

\section{Conclusion}

In summary, we studied the effects of perfect fluid dark matter and quantum corrections on the thermodynamics of nonlinear magnetic-charged black hole. First of all, we used the horizon propriety to find the black hole mass, and the surface gravity to find the temperature. From the plot of the temperature, we showed that it undergoes a phase of decrease after having increased and reached a maximum. Furthermore, we showed that perfect fluid dark matter increases the maximum of temperature.

Secondly, we found the corrected entropy due to thermal fluctuation. Analyzing its behavior revealed that thermal fluctuation impacts small size black holes, since the corrected entropy violates the second law of thermodynamics, leading a nonlinear evolution and a decrease of the entropy. However, we showed that thermal fluctuation does not have a great effect on the entropy for larger black holes.

Thirdly, in order to study the effects of dark matter and thermal fluctuation on the stability of black hole, we plotted the heat capacity and the Gibbs free energy. Hence, we showed that the black hole undergoes a second-order phase transition. Unless the phase transition appears at the same place with or without quantum corrections, we showed that the heat capacity is affected by thermal fluctuation for smaller black holes.  Also, the analysis of the Gibbs free energy showed that thermal fluctuation modifies its behavior for small horizon radii, and its behaviour is changed through the second order phase transition. 

\bibliographystyle{unsrt}
\bibliography{bibli2}

\end{document}